\documentclass[12pt]{article}
\usepackage{amssymb,amsmath,epsfig}

\begin{document}

\title{\bf Effects of Electromagnetic Field on Gravitational Collapse}

\author{M. Sharif \thanks{msharif@math.pu.edu.pk} and G. Abbas
\thanks{abbasghulam47@yahoo.com}\\
Department of Mathematics, University of the Punjab,\\
Quaid-e-Azam Campus, Lahore-54590, Pakistan.}

\date{}
\maketitle

\begin{abstract}
In this paper, the effect of electromagnetic field has been
investigated on the spherically symmetric collapse with the perfect
fluid in the presence of positive cosmological constant. Junction
conditions between the static exterior and non-static interior
spherically symmetric spacetimes are discussed. We study the
apparent horizons and their physical significance. It is found that
electromagnetic field reduces the bound of cosmological constant by
reducing the pressure and hence collapsing process is faster as
compared to the perfect fluid case. This work gives the
generalization of the perfect fluid case to the charged perfect
fluid. Results for the perfect fluid case are recovered.
\end{abstract}

{\bf Keywords}: Electromagnetic Field, Gravitational Collapse,
Cosmological Constant.\\\\
{\bf PACS numbers:} 04.20.-q, 04.40.Dg, 97.10.CV

\section{Introduction}

General Relativity (GR) predicts that gravitational collapse of
massive objects (having mass $=10^6M_\odot~-~10^8M_\odot$, where
$M_\odot$ is mass of the Sun) results to the formation of spacetime
singularities in our universe \cite{1}. The singularity theorems
\cite{2} of Hawking and Penrose reveal that if a trapped surface
forms during the collapse of compact object, such a collapse will
develop a spacetime singularity. According to these theorems, the
occurrence of spacetime singularity (can be observed or not) is the
generic property of the spacetime in GR. An observable singularity
is called \emph{naked singularity} while other is called \emph{black
hole or covered singularity}.

An open and un-resolved problem in GR is to determine the final fate
of the gravitational collapse (i.e., end product of collapse is
either covered or naked singularity). To resolve this problem,
Penrose \cite{3} suggested a hypothesis so-called Cosmic Censorship
Conjecture (CCC). This conjecture states that the singularities that
appear in the gravitational collapse are always covered by the event
horizon. It has two versions, i.e., weak and strong version
\cite{4}. The weak version states that the gravitational collapse
from the regular initial conditions never creates spacetime
singularity visible to distant observer. On the other hand, the
strong version says that no singularity is visible to any observer
at all, even some one close to it. There is no mathematical or
theoretical proof for either of the version of the CCC.

The singularity (at the end stage of the gravitational collapse) can
be black hole or naked depending upon the initial data and equation
of the state. To prove or disprove this hypothesis, many efforts
have been made but no final conclusion is drawn. It would be easier
to find the counter example that would enable us to claim that the
hypothesis is not correct. For this purpose, Virbhadra et al.
\cite{5} introduced a new theoretical tool using the gravitational
lensing phenomena. Also, Virbhadra and Ellis \cite{6} studied the
Schwarzschild black hole lensing and found that the relativistic
images would confirm the Schwarzschild geometry close to the event
horizon. The same authors \cite{7} analyzed the gravitational
lensing by a naked singularity and classified it into two kinds:
weak naked singularity (those contained within at least one photon
sphere) and strong naked singularity (those not contained within any
photon sphere).

Claudel et al. \cite{8} showed that spherically symmetric black
holes, with reasonable energy conditions, are always covered inside
at least one photon sphere. Virbhadra and Keeton \cite{9} studied
the time delay and magnification centroid due to gravitational
lensing by black hole and naked singularity. It was found that weak
CCC can be tested observationally without any ambiguity. Virbhadra
\cite{10} explored the useful results to investigate the Seifert's
conjecture for naked singularity. He found that naked singularity
forming in the Vaidya null dust collapse supports the Seifert's
conjecture. In a recent paper \cite{11}, the same author used the
gravitational lensing phenomena to find the improved form of the
CCC. This work is a source of inspiration for many leading
researchers.

Oppenheimer and Snyder \cite{12} are the pioneers who investigated
gravitational collapse long time ago in 1939. They studied the dust
collapse by taking the static Schwarzschild spacetime as exterior
and Friedmann like solution as interior spacetime. They found black
hole as end product of the gravitational collapse. To study the
gravitational collapse, exact solutions of the Einstein field
equations with dust provide non-trivial examples of naked
singularity formation. Since the effects of pressure cannot be
neglected in the singularity formation, therefore dust is not
assumed to be a good matter.

There has been a growing interest to study the gravitational
collapse in the presence of perfect fluid and other general physical
form of the fluid. Misner and Sharp \cite{13} extended the pioneer
work for the perfect fluid. Vaidya \cite{14} and Santos \cite{15}
used the idea of outgoing radiation of the collapsing body and also
included the dissipation in the source by allowing the radial heat
flow. The cosmological constant $\Lambda$ affects the properties of
spacetime as it appears in the field equations. It is worthwhile to
solve the field equations with non-zero cosmological constant for
analyzing the gravitational collapse. Markovic and Shapiro \cite{20}
generalized the pioneer work with positive cosmological constant.
Lake \cite{21} extended it for both positive and negative
cosmological constant.

Sharif and Ahmad \cite{22}-\cite{22c} extended the spherically
symmetric gravitational collapse with positive cosmological constant
for perfect fluid. They discussed the junction conditions, apparent
horizons and their physical significance. It is concluded that
apparent horizon forms earlier than singularity and positive
cosmological constant slows down the collapse. The same authors also
investigated the plane symmetric gravitational collapse using
junction conditions \cite{23}. In a recent paper \cite{24}, Sharif
and Iqbal extended plane symmetric gravitational collapse to
spherically symmetric case.

Although a lot of work has been done for dust and perfect fluid
collapse of spherically symmetric models. However, no such attempt
has been made by including the electromagnetic field. We would like
to study the gravitational collapse of charged perfect fluid in the
presence of positive cosmological constant. For this purpose, we
discuss the junction conditions between the non-static interior and
static exterior spherically symmetric spacetimes. The main
objectives of this work are the following:
\begin{itemize}
\item To study the effects of electromagnetic field on the rate of
collapse.
\item  To see whether or not CCC is valid in this framework.
\end{itemize}

The plan of the paper is as follows: In the next section, the
junction conditions are given. We discuss the solution of the
Einstein-Maxwell field equations in section \textbf{3}. The apparent
horizons and their physical significance are presented in section
\textbf{4}. We conclude our discussion in the last section.

We use the geometrized units (i.e., the gravitational constant $G$=1
and speed of light in vacuum $c$=1 so that $M\equiv\frac{MG}{c^2}$
and $\kappa\equiv\frac{8\pi G}{c^4}=8\pi$). All the Latin and Greek
indices vary from 0 to 3, otherwise it will be mentioned.

\section{Junction Conditions}

We consider a timelike $3D$ hypersurface $\Sigma$ which separates
two 4$D$ manifolds $M^-$ and $M^+$ respectively. For the interior
manifold $M^-$, we take spherically symmetric spacetime given by
\begin{equation}\label{1}
ds_-^2=dt^2-X^2dr^2-Y^2(d\theta^2+\sin\theta^2d\phi^2),
\end{equation}
where $X$ and $Y$ are functions of $t$ and $r$. For the exterior
manifold $M^+$, we take Reissner-Nordstr$\ddot{o}$m de-Sitter
spacetime
\begin{equation}\label{2}
ds_+^2=NdT^2-\frac{1}{N}dR^2-R^2(d\theta^2+\sin\theta^2d\phi^2),
\end{equation}
where
\begin{equation}\label{3}
N(R)=1-\frac{2M}{R}+\frac{Q^2}{R^2}-\frac{\Lambda}{3}R^2,
\end{equation}
$M$ and $\Lambda$ are constants and $Q$ is the charge. The Israel
junction conditions are the following \cite{25}:
\begin{enumerate}
\item The continuity of line element over $\Sigma$ gives
\begin{equation}\label{4}
(ds^2_-)_{\Sigma}=(ds^2_+)_{\Sigma}=ds^2_{\Sigma}.
\end{equation}
\item The continuity of extrinsic curvature over $\Sigma$
yields
\begin{equation}\label{5}
[K_{ij}]=K^+_{ij}-K^-_{ij}=0,\quad(i,j=0,2,3)
\end{equation}
where $K_{ij}$ is the extrinsic curvature defined as
\end{enumerate}
\begin{equation}\label{6}
K^{\pm}_{ij}=-n^{\pm}_{\sigma}(\frac{{\partial}^2x^{\sigma}_{\pm}}
{{\partial}{\xi}^i{\partial}{\xi}^j}+{\Gamma}^{\sigma}_{{\mu}{\nu}}
\frac{{{\partial}x^{\mu}_{\pm}}{{\partial}x^{\nu}_{\pm}}}
{{\partial}{\xi}^i{\partial}{\xi}^j}),\quad({\sigma},
{\mu},{\nu}=0,1,2,3).
\end{equation}
Here $\xi^i$ correspond to the coordinates on ${\Sigma }$,
$x^{\sigma}_{\pm}$ stand for coordinates in $M^{\pm}$,  the
Christoffel symbols $\Gamma^{\sigma}_{{\mu}{\nu}}$ are calculated
from the interior or exterior spacetimes and $n^{\pm}_{\sigma}$ are
components of outward unit normals to ${\Sigma}$ in the coordinates
$x^{\sigma}_{\pm}$.

The equation of hypersurface in terms of interior spacetime $M^-$
coordinates is
\begin{equation}\label{8}
f_-(r,t)=r-r_{\Sigma}=0,
\end{equation}
where $r_{\Sigma}$ is a constant. Also, the equation of hypersurface
in terms of exterior spacetime $M^+$ coordinates is given by
\begin{equation}\label{9}
f_+(R,T)=R-R_{\Sigma}(T)=0.
\end{equation}
When we make use of Eq.(\ref{8}) in Eq.(\ref{1}), the metric on
$\Sigma$ takes the form
\begin{equation}\label{10}
(ds_-^2)_\Sigma={dt^2-Y^2(r_\Sigma
,t)(d\theta^2+\sin\theta^2d\phi^2)}.
\end{equation}
Also, Eqs.(\ref{9}) and (\ref{2}) yield
\begin{equation}\label{11}
(ds_+^2)_\Sigma=[N(R_\Sigma)-\frac{1}{N(R_\Sigma)}
(\frac{dR_\Sigma}{dT})^2]dT^2-R_\Sigma^2(d\theta^2+\sin\theta^2d\phi^2),
\end{equation}
where we assume that
\begin{equation}\label{12}
N(R_\Sigma)-\frac{1}{N(R_\Sigma)} (\frac{dR_\Sigma}{dT})^2>0
\end{equation}
so that T is a timelike coordinate. From Eqs.(\ref{4}), (\ref{10})
and (\ref{11}), it follows that
\begin{eqnarray}\label{13}
R_\Sigma=Y(r_\Sigma,t),\\\label{14}
[N(R_\Sigma)-\frac{1}{N(R_\Sigma)}
(\frac{dR_\Sigma}{dT})^2]^{\frac{1}{2}}dT=dt .
\end{eqnarray}
Also, from Eqs.(\ref{8}) and (\ref{9}), the outward unit normals in
$M^-$ and $M^+$, respectively, are given by
\begin{eqnarray}\label{15}
n^-_\mu&=&(0,X(r_\Sigma,t),0,0),\\
\label{16} n^+_\mu&=&(-\dot{R}_\Sigma,\dot{T}, 0,0).
\end{eqnarray}
The components of extrinsic curvature $K^\pm_{ij}$ become
\begin{eqnarray}\label{17}
K^-_{00}&=&0,\\
\label{18}
K^-_{22}&=&\csc^2{\theta}K^-_{33}=(\frac{YY'}{X})_\Sigma,\\
\label{19}
K^+_{00}&=&(\dot{R}\ddot{T}-\dot{T}\ddot{R}-\frac{N}{2}\frac{dN}{dR}\dot{T}^3
+\frac{3}{2N}\frac{dN}{dR}\dot{T}\dot{R}^2)_\Sigma,\\
\label{20} K^+_{22}&=&\csc^2{\theta}
K^+_{33}=(NR\dot{T})_{\Sigma},
\end{eqnarray}
where dot and prime mean differentiation with respect to $t$ and $r$
respectively. From Eq.(\ref{5}), the continuity of extrinsic
curvature gives
\begin{eqnarray}\label{21}
K^+_{00}&=&0,\\
\label{22} K^+_{22}&=&K^-_{22}.
\end{eqnarray}
Using Eqs.(\ref{17})-(\ref{22}) along with Eqs.(\ref{3}), (\ref{13})
and (\ref{14}), the junction conditions become
\begin{eqnarray}\label{23}
(X\dot{{Y'}}-\dot{X}{Y'})_\Sigma=0,\\
\label{24}
M=(\frac{Y}{2}-\frac{\Lambda}{6}Y^3+\frac{Q^2}{2Y}
+\frac{Y}{2}\dot{Y}^2-\frac{Y}{2X^2}{Y'}^2)_{\Sigma}.
\end{eqnarray}

\section{Solution of the Einstein Field Equations}

The Einstein field equations with cosmological constant are given by
${\setcounter{equation}{0}}$
\begin{equation}\label{25}
G_{\mu\nu}-{\Lambda}g_{\mu\nu}=\kappa(T_{\mu\nu}+T^{({em})}_{\mu\nu}).
\end{equation}
The energy-momentum tensor for perfect fluid is
\begin{equation}\label{26}
{T_{{\mu}{\nu}}={({\rho}+p)}u_{\mu}u_{\nu}-pg_{\mu\nu}},
\end{equation}
where $\rho$ is the energy density, $p$ is the pressure and
$u_\mu=\delta^0_\mu$ is the four-vector velocity in co-moving
coordinates. $T^{({em})}_{\mu\nu}$ is the energy-momentum tensor for
the electromagnetic field defined \cite{26} as
\begin{equation}\label{27}
T^{(em)}_{{\mu}{\nu}}=\frac{1}{4{\pi}}(-g^{{\delta}{\omega}}
F_{{\mu}{\delta}}F_{{\nu}{\omega}}+\frac{1}{4}g_{{\mu}{\nu}}
F_{{\delta}{\omega}}F^{{\delta}{\omega}}).
\end{equation}
With the help of Eqs.(\ref{26}) and (\ref{27}), Eq.(\ref{25}) takes
the form
\begin{equation}\label{28}
R_{{\mu}{\nu}}=8\pi[({\rho}+p)u_{\mu}u_{\nu}
+\frac{1}{2}(p-{\rho})g_{{\mu}{\nu}}
+T^{({em})}_{{\mu}{\nu}}-\frac{1}{2}g_{{\mu}{\nu}}T^{({em})}]
-{\Lambda}g_{{\mu}{\nu}}.
\end{equation}
To solve this equation, we need to calculate the non-zero components
and trace free form of $T^{({em})}_{{\mu}{\nu}}$. For this purpose,
we first solve the Maxwell's field equations
\begin{eqnarray}\label{29}
F_{\mu\nu}&=&\phi_{\nu,\mu}-\phi_{\mu,\nu},\\\label{30}
F^{\mu\nu}_{}{;\nu}&=&-4{\pi}J^{\mu},
\end{eqnarray}
where $\phi_{\mu}$ is the four potential and $J^{\mu}$ is the four
current. As the charged fluid is in co-moving coordinate system, the
magnetic field will be zero in this case. Thus we can choose the
four potential and four current as follows
\begin{eqnarray}\label{31}
\phi_{\mu}&=&({\phi}(t,r),0,0,0),\\\label{32}
J^{\mu}&=&{\sigma}u^{\mu},
\end{eqnarray}
where $\sigma$ is charge density.

Now for the solution of the Maxwell's field Eq.(\ref{30}), $\mu$ and
$\nu$ are treated as local coordinates. Using Eqs.(\ref{29}) and
(\ref{31}), the non-zero components of the field tensor are given as
follows:
\begin{equation}\label{33}
F_{{0}{1}}=-F_{{1}{0}}=-\frac{{\partial}{\phi}}{{\partial}{r}}.
\end{equation}
Also, from Eqs.(\ref{30}) and (\ref{32}), we have
\begin{equation}\label{34}
\frac{1}{X}\frac{{\partial}^2{\phi}}{{\partial}{r}^2}-\frac{{\partial}{\phi}}
{{\partial}{r}}\frac{X'}{X^2}=-4{\pi}{\sigma}X,
\end{equation}
\begin{equation}\label{35}
\frac{1}{X}\frac{{\partial}^2{\phi}}{{\partial}{r}{\partial}{t}}
-\frac{\dot{X}}{X^2}\frac{{\partial}{\phi}}{{\partial}{r}}=0.
\end{equation}
Equation (\ref{35}) implies that
\begin{equation}\label{36}
(\frac{1}{X}\frac{{\partial}{\phi}}{{\partial}{r}})=K,
\end{equation}
where $K=K(r)$ is an arbitrary function of $r$. Equations (\ref{34})
and (\ref{36}) yield
\begin{equation}\label{37}
K'(r)=-4{\pi}{\sigma}X.
\end{equation}
The non-zero components of $T^{(em)}_{{\mu}{\nu}}$ and its trace
free form turn out to be
\begin{eqnarray*}
T^{(em)}_{{0}{0}}&=&\frac{1}{8{\pi}}K^2 ,\quad
T^{(em)}_{{1}{1}}=-\frac{1}{8{\pi}}K^2X^2 ,\quad
T^{(em)}_{{2}{2}}=\frac{1}{8{\pi}}K^2Y^2,\\
T^{(em)}_{{3}{3}}&=&T^{(em)}_{{2}{2}}\sin^2\theta,\quad
T^{(em)}=0.
\end{eqnarray*}

When we use these values, the field equations (\ref{28}) for the
interior spacetime takes the form
\begin{eqnarray}\label{42}
R_{00}&=&-\frac{\ddot{X}}{X}-2\frac{\ddot{Y}}{Y}=4\pi(\rho+3p)
+K^2-{\Lambda},\\
\label{43}
R_{11}&=&-\frac{\ddot{X}}{X}-2\frac{\dot{X}}{X}\frac{\dot{Y}}{Y}
+\frac{2}{X^2}[\frac{Y''}{Y}-\frac{Y'X'}{XY}]\nonumber\\
&=&{4\pi}(p-\rho)+K^2-{\Lambda} ,\\
\label{44}
R_{22}&=&-\frac{\ddot{Y}}{Y}-(\frac{\dot{Y}}{Y})^2
-\frac{\dot{X}}{X}\frac{\dot{Y}}{Y}
+\frac{2}{X^2}[\frac{Y''}{Y}+(\frac{Y'}{Y})^2-
\frac{X'}{X}\frac{Y'}{Y}-(\frac{X}{Y})^2]\nonumber\\
&=&{4\pi}(p-\rho)-K^2-{\Lambda} ,\\
\label{45} R_{33}&=&{\sin}^2{\theta}R_{22} ,\\
\label{46}
R_{01}&=&-2\frac{\dot{Y'}}{Y}+2\frac{\dot{X}}{X}\frac{Y'}{Y}=0 .
\end{eqnarray}
Now we solve Eqs.(\ref{42})-(\ref{46}). Integration of Eq.(\ref{46})
with respect to $t$ yields
\begin{equation}\label{47}
X=\frac{Y'}{H},
\end{equation}
where $H=H(r)$ is an arbitrary function of $r$. The energy
conservation equation
\begin{equation}\label{48}
T^{\nu}_{{\mu};{\nu}}=0,
\end{equation}
for the perfect fluid with the interior metric shows that pressure
is a function of $t$ only, i.e.,
\begin{equation}\label{49}
p=p(t).
\end{equation}

Substituting the values of $X$ and $p$ from Eqs.(\ref{47}) and
(\ref{49}) in Eqs.(\ref{42})-(\ref{46}), it follows that
\begin{equation}\label{50}
2\frac{\ddot{Y}}{Y}+(\frac{\dot{Y}}{Y})^2+\frac{(1-H^2)}{Y^2}
=\Lambda+{K^2}-p(t).
\end{equation}
We consider $p$ as a polynomial in $t$ as given by \cite{22a}
\begin{equation}\label{51}
p(t)=p_ct^{-s},
\end{equation}
where $p_c$ and $s$ are positive constants. Further, for simplicity,
we take $s=0$ so that
\begin{equation}\label{52}
p(t)=p_c.
\end{equation}
Replacing this value in Eq.(\ref{50}), we get
\begin{equation}\label{53}
2\frac{\ddot{Y}}{Y}+(\frac{\dot{Y}}{Y})^2+\frac{(1-H^2)}{Y^2}
=\Lambda+{K^2}-{8\pi}p_c.
\end{equation}
Integrating this equation with respect to $t$, it follows that
\begin{equation}\label{54}
{\dot{Y}}^2=H^2-1+(\Lambda+{K^2}-{8\pi}p_c)\frac{Y^2}{3}+2\frac{m}{Y},
\end{equation}
where $m=m(r)$ is an arbitrary function of $r$ and is related to the
mass of the collapsing system. Substituting Eqs.(\ref{47}),
(\ref{54}) into Eq.(\ref{42}), we get
\begin{equation}\label{55}
m'=\frac{2K'K}{3}Y^3+{{Y'}{Y^2}}[4\pi(p_c+{\rho})+2{K^2}].
\end{equation}

For physical reasons, we assume that pressure and density are
strictly positive. Integrating Eq.(\ref{55}) with respect to $r$, we
obtain
\begin{equation}\label{56}
m(r)=4\pi\int^r_0({\rho}+{p_c}){Y'}{Y^2}dr+2\int^r_0
K^2{Y'}{Y^2}dr+\frac{2}{3}\int^r_0{K'K}Y^3dr.
\end{equation}
The function $m(r)$ must be positive because $m(r)<0$ implies
negative mass which is not physical. Using Eqs.(\ref{47}) and
(\ref{54}) into the junction condition Eq.(\ref{24}), it follows
that
\begin{equation}\label{57}
M=\frac{Q^2}{2Y}+m+\frac{1}{6}(\Lambda+K^2-{8\pi}p_c)Y^3.
\end{equation}
The total energy $\tilde{M}(r,t)$ up to a radius $r$ at time $t$
inside the hypersurface $\Sigma$ can be evaluated by using the
definition of mass function \cite{13} given by
\begin{equation}\label{58}
\tilde{M}(r,t)=\frac{1}{2}Y(1+g^{\mu\nu}Y_{,\mu} Y_{,\nu}).
\end{equation}
For the interior metric, it takes the form
\begin{equation}\label{59}
\tilde{M}(r,t)=\frac{1}{2}Y(1+\dot{Y}^2-(\frac{Y'}{X})^2).
\end{equation}
Replacing Eqs.(\ref{47}) and (\ref{54}) in Eq.(\ref{59}), we obtain
\begin{equation}\label{60}
\tilde{M}(r,t)=m(r)+(\Lambda+{K^2}-{8\pi}p_c)\frac{Y^3}{6}.
\end{equation}

Now we take $(\Lambda+{K^2}-{8\pi}p_c)>0$ and the assumption
\begin{equation}\label{61}
H(r)=1.
\end{equation}
In order to obtain the analytic solutions in closed form, we use
Eqs.(\ref{47}), (\ref{54}) and (\ref{61}) so that
\begin{eqnarray}\label{62}
Y&=&(\frac{6m}{\Lambda+{K^2}-{8\pi}p_c})^\frac{1}{3}\sinh\alpha(r,t),\\
\label{63}
X&=&(\frac{6m}{\Lambda+{K^2}-{8\pi}p_c})^\frac{1}{3}\left.[\{\frac{m'}{3m}-\frac{2K
K'}{3(\Lambda+{K^2}-{8\pi}p_c)}\}\sinh\alpha(r,t)\right.\nonumber\\
&+&\left.\{\frac{2(t_s(r)-t){KK'}}{\sqrt{3(\Lambda+{K^2}-{8\pi}p_c)}}
+t'_s(r)\sqrt{\frac{\Lambda+{K^2}-{8\pi}p_c}{3}}\}\right.\nonumber\\
&\times&\left.\cosh\alpha(r,t)\right.]\sinh^\frac{-1}{3}\alpha(r,t),
\end{eqnarray}
where
\begin{equation}\label{64}
\alpha(r,t)=\frac{\sqrt{3(\Lambda+{K^2}-{8\pi}p_c)}}{2}[t_s(r)-t)].
\end{equation}
Here $t_s(r)$ is an arbitrary function of $r$ and is related to the
time of formation of singularity of a particular shell at coordinate
distance $r$.

In the limit $({8\pi}p_c-{K^2})\rightarrow\Lambda$, the above
solution corresponds to the Tolman-Bondi solution \cite{27}
\begin{eqnarray}\label{65}
\lim_{({8\pi}p_c-{K^2})\longrightarrow\Lambda}X(r,t)&=&\frac{m'(t_s-t)+2mt'_s}
{[6m^2(t_s-t)]^\frac{1}{3}},\\
\label{66} \lim_{({8\pi}p_c-{K^2})\longrightarrow\Lambda}Y(r,t)&=&
[\frac{9m}{2}(t_s-t)^2]^{\frac{1}{3}}.
\end{eqnarray}

\section{Apparent Horizons}

Here we discuss the apparent horizons for the interior spacetime.
The boundary of two trapped spheres whose outward normals are null
is used to find the apparent horizons. This is given as follows:
${\setcounter{equation}{0}}$
\begin{equation}\label{67}
g^{\mu\nu}Y_{,\mu} Y_{,\nu}=\dot{Y}^2-(\frac{Y'}{X})^2=0.
\end{equation}
Replacing Eqs.(\ref{47}) and (\ref{54}) in this equation, we get
\begin{equation}\label{68}
(\Lambda+{K^2}-{8\pi}p_c)\frac{Y^3}{3}-3Y+6m=0.
\end{equation}
When we take $\Lambda=8\pi p_c-{K^2}$, it gives $Y=2m$. This is
called Schwarzschild horizon. For $m=p_c=K=0$, we have
$Y=\sqrt{\frac{3}{\Lambda}}$, which is called de-Sitter horizon.
Equation (\ref{68}) can have the following positive roots.\\\\
\textbf{Case (i)}: For
$3m<\frac{1}{\sqrt{(\Lambda+{K^2}-{8\pi}p_c)}}$, we obtain two
horizons
\begin{eqnarray}\label{69}
Y_1&=&\frac{2}{\sqrt{(\Lambda+{K^2}-{8\pi}p_c)}}\cos\frac{\varphi}{3},\\
\label{70}
Y_2&=&\frac{-1}{\sqrt{(\Lambda+{8\pi}{K^2}-p_c)}}
(\cos\frac{\varphi}{3}-\sqrt{3}\sin\frac{\varphi}{3}),
\end{eqnarray}
where
\begin{equation}\label{71}
\cos\varphi=-3m{\sqrt{(\Lambda+{K^2}-{8\pi}p_c)}}.
\end{equation}
If we take $m=0$, it follows from Eqs.(\ref{69}) and (\ref{70}) that
$Y_1=\sqrt{\frac{3}{(\Lambda+{K^2}-{8\pi}p_c)}}$ and $Y_2=0$. $Y_1$
and $Y_2$ are called cosmological horizon and black hole horizon
respectively. For $m\neq0$ and $\Lambda\neq{8\pi}p_c-{K^2}$, $Y_1$
and $Y_2$ can be generalized \cite{28} respectively.\\\\
\textbf{Case (ii):} For
$3m=\frac{1}{\sqrt{(\Lambda+{K^2}-{8\pi}p_c)}}$, there is only one
positive root which corresponds to a single horizon i.e.,
\begin{equation}\label{72}
Y_1=Y_2=\frac{1}{\sqrt{(\Lambda+{K^2}-{8\pi}p_c)}}=Y.
\end{equation}
This shows that both horizons coincide. The range for the
cosmological and black hole horizon can be written as follows
\begin{equation}\label{73}
0\leq Y_{2} \leq \frac{1}{\sqrt{(\Lambda+{K^2}-{8\pi}p_c)}} \leq
Y_{1} \leq \sqrt{\frac{3}{(\Lambda+{K^2}-{8\pi}p_c)}} .
\end{equation}
The black hole horizon has its largest proper area
${4\pi}Y^2=\frac{4\pi}{(\Lambda+{K^2}-{8\pi}p_c)}$ and cosmological
horizon has its area between
$\frac{4\pi}{(\Lambda+{K^2}-{8\pi}p_c)}$and
$\frac{12\pi}{(\Lambda+{K^2}-{8\pi}p_c)}$.\\\\
\textbf{Case (iii):} For
$3m>\frac{1}{\sqrt{(\Lambda+{K^2}-{8\pi}p_c)}}$, there are no
positive roots and consequently there are no apparent horizons.

We now calculate the time of formation for the apparent horizon
using Eqs.(\ref{61}), (\ref{62}) and (\ref{68})
\begin{equation}\label{74}
t_n=t_s-\frac{2}{\sqrt{3(\Lambda+{K^2}-{8\pi}p_c})}\sinh^{-1}
(\frac{Y_n}{2m}-1)^{\frac{1}{2}},
\quad(n=1,2).
\end{equation}
When ${8\pi}p_c-{K^2}\longrightarrow\Lambda$, this corresponds to
Tolman-Bondi \cite{27}
\begin{equation}\label{75}
t_{ah}=t_s-\frac{4}{3}m.
\end{equation}
From Eq.(\ref{74}), we can write
\begin{equation}\label{76}
\frac{Y_n}{2m}=\cosh^{2}\alpha_n,
\end{equation}
where
$\alpha_n(r,t)=\frac{\sqrt{3(\Lambda+{K^2}-{8\pi}p_c)}}{2}[t_s(r)-t_n)]$.
Equations (\ref{73}) and (\ref{74}) imply that $Y_{1}\geq Y_{2}$ and
$t_{2} \geq t_{1}$ respectively. The inequality $t_{2} \geq t_{1}$
indicates that the cosmological horizon forms earlier than the black
hole horizon.

The time difference between the formation of cosmological horizon
and singularity and the formation of black hole horizon and
singularity respectively can be found as follows. Using
Eqs.(\ref{69})-(\ref{71}), it follows that
\begin{eqnarray}\label{77}
\frac{d(\frac{Y_1}{2m})}{dm}&=&\frac{1}{m}(-\frac{\sin\frac{\varphi}{3}}{\sin\varphi}
+\frac{3\cos\frac{\varphi}{3}}{\cos\varphi})<0,\\
\label{78}\frac{d(\frac{Y_2}{2m})}{dm}
&=&\frac{1}{m}(-\frac{\sin\frac{(\varphi+4\pi)}{3}}{\sin\varphi}
+\frac{3\cos\frac{(\varphi+4\pi)}{3}}{\cos\varphi})>0.
\end{eqnarray}
The time difference between the formation of singularity and
apparent horizons is
\begin{equation}\label{79}
T_n=t_s-t_n.
\end{equation}
It follows from Eq.(\ref{76}) that
\begin{equation}\label{80}
\frac{dT_n}{d(\frac{Y_n}{2m})}
=\frac{1}{\sinh\alpha_n\cosh\alpha_n{\sqrt{3(\Lambda+{K^2}-{8\pi}p_c)}}}.
\end{equation}
Using Eqs.(\ref{77}) and (\ref{80}), we get
\begin{eqnarray}\label{81}
\frac{dT_1}{dm}=\frac{dT_1}{d(\frac{Y_1}{2m})}\frac{d(\frac{Y_1}{2m})}{dm}
=\frac{1}{m{\sqrt{3(\Lambda+{K^2}-{8\pi}p_c)}}\sinh\alpha_1\cosh\alpha_1}\nonumber\\
\times(-\frac{\sin\frac{\varphi}{3}}{\sin\varphi}
+\frac{3\cos\frac{\varphi}{3}}{\cos\varphi})<0 .
\end{eqnarray}
It shows that $T_1$ is a decreasing function of mass $m$. This means
that time interval between the formation of cosmological horizon and
singularity is decreased with the increase of mass. Similarly, from
Eqs.(\ref{78}) and(\ref{80}), we get
\begin{eqnarray}\label{82}
\frac{dT_2}{dm}= \frac{1}{m{\sqrt{3
(\Lambda+{K^2}-{8\pi}p_c)}}\sinh\alpha_2\cosh\alpha_2}\nonumber\\
\times(-\frac{\sin\frac{(\varphi+4\pi)}{3}}{\sin\varphi}
+\frac{3\cos\frac{(\varphi+4\pi)}{3}}{\cos\varphi})>0.
\end{eqnarray}
This indicates that $T_2$ is an increasing function of mass $m$
indicating that time difference between the formation of black hole
horizon and singularity is increased with the increase of mass.

\section{Summary and Conclusion}

This paper is devoted to study the effects of electromagnetic field
on gravitational collapse with the positive cosmological constant.
The cosmological constant acts as Newtonian potential. The relation
for the Newtonian potential is $\phi=\frac{1}{2}(1-g_{00})$. Using
Eqs.(\ref{13}) and (\ref{57}), for the exterior spacetime, the
Newtonian potential turns out to be ${\setcounter{equation}{0}}$
\begin{equation}\label{84}
\phi(R)=\frac{m}{R}+(\Lambda+{K^2}-{8\pi}p_c)\frac{R^2}{6}.
\end{equation}
The corresponding Newtonian force is
\begin{equation}\label{85}
F=-\frac{m}{R^2}+(\Lambda+{K^2}-{8\pi}p_c)\frac{R}{3} .
\end{equation}

Now we discuss the consequence of the Newtonian force. This force is
zero for the fixed values of
$m=\frac{1}{3\sqrt{(\Lambda+{K^2}-{8\pi}p_c)}}$ and
$R=\frac{1}{\sqrt{(\Lambda+K^2-{8\pi}p_c)}}$ and will be positive
(repulsive) if the values of $m$ and $R$ are taken larger than these
values. If we take $m=\frac{1}{\sqrt{(\Lambda+{K^2}-{8\pi}p_c)}}$
and $R=\frac{3}{\sqrt{(\Lambda+K^2-{8\pi}p_c)}}$, then
$F=\frac{2(\Lambda+{K^2}-{8\pi}p_c)}{9}$ which gives positive value
if $(\Lambda+{K^2}-{8\pi}p_c)>0$, i.e., $\Lambda
>({8\pi}p_c-{K^2})$ such that ${8\pi}p_c>{K^2}$.
Thus we conclude that the repulsive force can be generated from
$\Lambda$ if $\Lambda >({8\pi}p_c-{K^2})$ such that
${8\pi}p_c>{K^2}$ over the entire range of the collapsing sphere.
For the perfect fluid and dust cases, $\Lambda$ can play the role of
the repulsive force for $\Lambda
>{8\pi}p_c$ and $\Lambda>0$ respectively. Notice that $K=K(r)$
gives the electromagnetic field contribution. From Eq.(\ref{54}),
the rate of collapse turns out be
\begin{equation}\label{86}
\ddot{Y}=-\frac{m}{Y^2}+(\Lambda+{K^2}-{8\pi}p_c)\frac{Y}{3}.
\end{equation}
This shows that we have re-formulated the Newtonian model which
represents the acceleration of the collapsing process. The analysis
of positive and negative acceleration would give the same results as
for the Newtonian force.

It is worthwhile to mention that the electromagnetic field reduces
the bound of the positive cosmological constant by reducing the
pressure. Thus the positive cosmological constant is bounded below
as compared to the perfect fluid case. This would decrease the
repulsive force which slows down the collapsing process. Making the
analysis of the smaller values of $m$ and $R$ than the values used
for the repulsive force, we find that the attractive force is larger
than the perfect fluid case. Since the attractive force favors the
collapse while the repulsive force resists against the collapse,
thus the collapsing process is faster as compared to perfect fluid
case when we include the electromagnetic field.

Further, we have found two apparent horizons (cosmological and black
hole horizons) whose area decreases in the presence of
electromagnetic field. It is found that the cosmological horizon
forms earlier than the black hole horizon. Also, Eq.(\ref{74}) shows
that apparent horizon forms earlier than singularity. In this sense,
we can conclude that the end state of gravitational collapse is a
singularity covered by the apparent horizons (i.e., black hole).

It is interesting to mention here that our study supports the CCC
and would be considered as one of its counter example. Also, it
would be possible that the electromagnetic field reduces the range
of apparent horizons to extreme limits and singularity would be
locally naked. Thus the weak version of the CCC seems to be valid in
this case.

\vspace{0.25cm}

{\bf Acknowledgment}

\vspace{0.25cm}

We would like to thank the Higher Education Commission, Islamabad,
Pakistan for its financial support through the {\it Indigenous Ph.D.
5000 Fellowship Program Batch-IV}.

\end{document}